# A case study: the savings potential thanks to FAIR data in one Materials Science PhD project

Michael Seitz, Nick Garabedian, Ilia Bagov, Christian Greiner

*Institute for Applied Materials (IAM), Karlsruhe Institute of Technology (KIT), Kaiserstr. 12, 76131 Karlsruhe, Germany*

*IAM-ZM MicroTribology Center (µTC), Strasse am Forum 5, 76131 Karlsruhe, Germany*

The FAIR (Findable, Accessible, Interoperable, and Reusable) data principles have gained significant attention as a means to enhance data sharing, collaboration, and reuse across various domains. Here, we explore the potential benefits of implementing FAIR data practices within engineering projects, with a monetary focus in the German context, but by considering aspects which are relatively universal. By examining the FAIR-data aspect of a Materials Science and Engineering PhD project, it becomes evident that substantial cost savings can be achieved. The estimated savings are € 2,600 per year from the PhD project considered.

This study underscores the importance of implementing FAIR data practices in engineering projects and highlights some significant economic benefits that can be derived from such initiatives. By embracing FAIR principles, organizations in the engineering sector can unlock the full potential of their data, optimize resource allocation, and drive innovation in a cost-effective manner.

1. Introduction

Data management needs to play a key role in today's scientific projects. On the one hand, larger data storage capabilities offer the possibility to record an increasing number of data sources at higher temporal resolution. On the other hand, data can be retrieved via the internet from anywhere in the world. In order to use the raw data obtained in experiments and simulations as efficiently as possible, detailed descriptions and explanations – so-called metadata – are essential. Efficiency in this context implies findability, accessibility, interoperability and reusability, which are best summarized under the acronym FAIR [1]. However, the digital transformation of individual laboratories and departments up to entire scientific communities and companies, aiming to implement FAIR data practices, is both time-consuming and cost-intensive [2].

Although setting up a large-scale data infrastructure might seem like the central problem standing in the way of a FAIR-er future, in fact, digitalization starts with a data culture change. Existing laboratories need to abandon "paper and flash drive" practices in favor of metadata-driven ones, and staff needs to be made aware of the new data concept. Previous studies focus on the macroscale costs of non-FAIR data (in 2019 it cost the European economy an estimated 10.2 billion € annually) [3] and are not informative at an individual lab level. For this reason, we here are looking at a typical PhD project (duration 3 years, full-time research assistant, 45 % technician) in the field of materials science and mechanical engineering to determine the monetary value of time lost that could have been avoided by applying the FAIR data principles. Subsequently, real examples are given to show how financial and time resources can be saved by applying FAIR data practices in everyday laboratory operations, in evaluating programming scripts, and also in scientific exchanges.

It should be noted that this report was evaluated for this particular project. In principle, however, it can be applied to other projects, even though structures and costs of different scientific projects vary depending on the country, research area, duration, hourly wage and many other factors. In this report, the following per-hour rates are applied, based on DFG (German Research Foundation) rates, shown in Table 1 (including overhead):

**Table 1**: Typical hourly rates for staff used in this report.

| JOB TITLE | SALARY | JOB DESCRIPTION |
| --- | --- | --- |
| **Research associate (PhD)** | 43.7 €/h (DFG rate, 40-hour week, 22 % overhead, can vary, based on 2022 data) | Setup of the experiment, project management, oversight, data analysis, simulations, presentation |
| **Research assistant (non-PhD student)** | 15.3 €/h | Experimental procedure, specimen preparation |
| **Technician** | 32.2 €/h | Specimen preparation, machine maintenance |

 2. **Selected PhD Project: Composite Peening**

In order to get a tangible impression of the financial loss due to non-FAIR data, a successfully-completed PhD project [4] in the field of materials science was examined in more detail. This project in "composite peening" was completed by Dr. Michael Seitz between 2016 and 2020 at Karlsruhe Institute of Technology's (KIT) Institute for Applied Materials (IAM). Composite peening is a novel process based on micro peening. By adding a heating device to a micro peening system, it is possible to introduce ceramic blasting particles several micrometers into the base material of aluminum, for example. The penetration depth depends on different process parameters such as materials used, temperature, peening pressure and number of treatments. Subsequently, the microstructure and mechanical properties were examined after composite peening. These include SEM and TEM analyses, as well as bending tests and investigations of the corrosive and tribological properties. Composite peening is described in detail in [4, 5].

It should be emphasized that this particular PhD project was completed successfully. However, a more efficient approach may have been possible with a holistic application of the FAIR data principles. Detailed examples are given below. Additionally, it has to be noted that the calculations in this case study are deliberately kept conservative.

Figure 1 shows a small part of the complexity and dependencies in this project. This figure is a reconstruction of the basic vocabulary used in the PhD project, and full resolution and machine operable versions are available at [6]. Unfortunately, the relationships between entities in a PhD project are rarely explicitly documented, and the best source of information become the associated journal publications (if existent), theses, and the lab notebook notes. Figure 1 shows only one of the many facets and ways of organizing the project's structure; for example, non-hierarchical relationships like the one between roughness measurements, surface topography, and solid particle erosion are not shown in the image. It quickly becomes apparent that there are many factors which have an influence on the structure of the generated surface modification and subsequently on the specimen's mechanical properties. Since such relationships are not obvious in the beginning of larger projects, a timely focused effort of FAIR data collection can be later crucial for its success, as they may be hidden otherwise.

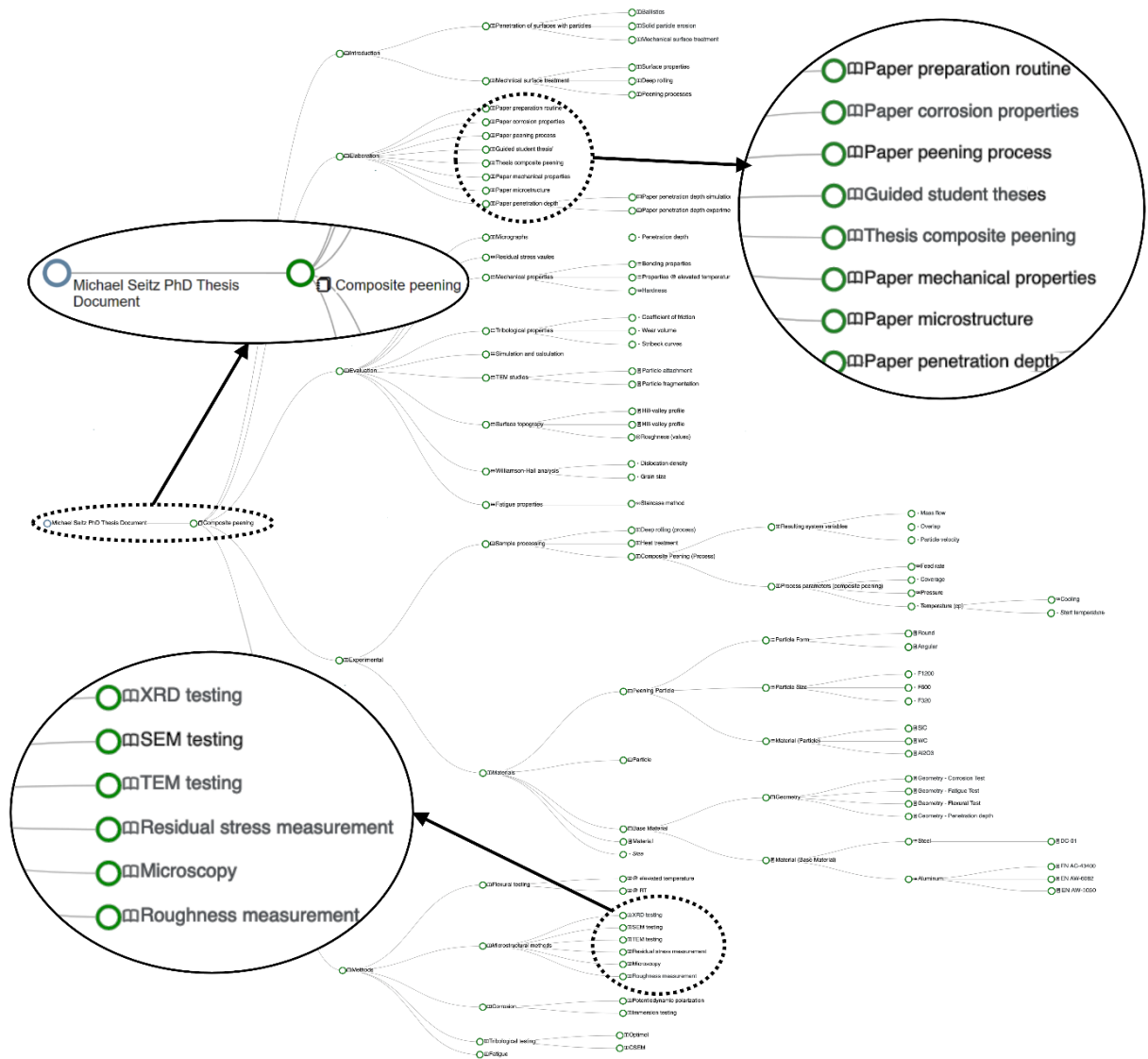

**Figure 1:** Structural composition of the PhD project – a hierarchical representation of the project structure. Diagram generated with the help of the metadata management software VocPopuli [7, 8]. This visual reconstruction is available in full resolution at [6].

To show the capability of FAIR data, four different elements of doctoral activities and the corresponding savings potential were considered. First, literature work supported by FAIR data is addressed. The highest missed financial benefit that FAIR data could have brought was found to be during lab work and data analysis in the two subsequent phases. Finally, publishing with the support of FAIR data is analyzed.

### 2.1. Phase 1: Literature Review and FAIR Data

By applying the FAIR data principles, the reuse of already generated data can be improved and accelerated. For better visibility, in most journal publications, data are presented graphically.

In the event that specific data behind a plot are needed for simulations or comparative experiments, raw data are most often not available or need to be manually requested via email from the authors of the manuscript in question. To illustrate the issue, a random issue of a prestigious materials science journal was selected (Journal of Materials Processing Technology, Volume 311, January 2023); the journal could be a place for a peening publication. Table 2 summarizes the types of statements listed under each publication in the selected volume. It is not encouraging that there is not a single paper that does not include a degree of difficulty when it comes to reusing the data. Alternatively, researchers often rely on programs that can estimate data from graphs, but this approach is time consuming and is also one additional source of uncertainty.

**Table 2**: A summary of Data Availability Statements from Journal of Materials Processing Technology, Volume 311, January 2023. No paper includes a permissionless way to obtain the raw data presented.

| DATA AVAILAIBILITY STATEMENT | COUNT |
|---|---|
| *Data will be made available on request.*<br>issue: In this case data can already be made public during publication. | 14 |
| *No data was used for the research described in the article.*<br>issue: In fact, all papers include graphs or micrographs. | 7 |
| *-No statement-*<br>issue: All mentioned issues in this table apply. | 3 |
| *The authors do not have permission to share data.*<br>issue: Without further information, it is hard to asses this statement. | 2 |
| *The data that has been used is confidential.*<br>issue: Without further information, it is hard to asses this statement. | 2 |
| *The raw/processed data required to reproduce these findings cannot be shared at this time due to technical or time limitations.*<br>issue: There is an expectation that peer-reviewed publications are verifiably reproducible. | 1 |
| *The raw/processed data required to reproduce these findings cannot be shared at this time as the data is currently being used in another study.*<br>issue: There must be a clear timeline for the publication of the data after the end of the other study. | 1 |
| *The authors are unable or have chosen not to specify which data has been used.*<br>issue: Authors are supposed to know what data is included in their study. | 1 |
| *Data is available for download at …* | 0 |

Relating this problem to the project at hand, in the composite peening project, 16 diagrams from literature were captured manually. Assuming a workload of half an hour per diagram, this results in **350 €** in potential savings at the hourly rate of a PhD candidate.

Even with a systematic peer review process, minor or major errors and inaccuracies can occur. This was recently discussed in terms of the motivations underlying the scientific publication of articles that include x-ray photoelectron spectroscopy (XPS) [9] – 40% of the (XPS) fits are incorrect, while another 40% are potentially incorrect. In the case of conducting a composite peening project, the problem that caused the most ambiguity was that instead of a scale bar, a magnification was provided for microscopic images. Since the magnification is not unambiguous due to different sized screens, printouts, etc., further use by other scientists is no longer possible without descriptive metadata. The same applies to experiments whose description has been formulated in a vague manner. For example, the humidity or room temperature may not seem relevant to the actual experiment, but this limits the reusability of the data. Often, machine parameters such as controller settings are also not fully specified.

### 2.2. Phase 2: FAIR Data in Lab Work

Novel procedures in science often inevitably lead to the repetition of experimental series, since it is not yet known where the focus of interest might be. In composite peening, for example, sample mix-up and incorrectly performed preparation steps lead to less valuable and even non-usable experimental data. Additionally, insufficiently recorded metadata (like humidity, room temperature) leads to a lack of knowledge retention and depth, which results in a time-consuming and expensive repetition of experiments. For the field of tribology, a recent example [10, 11] showcases a framework for recording detailed domain information, which would often be impossible to record in a paper lab note book because of the low bandwidth of written text compared to digitalized descriptions. This approach aims to counteract cases where, for example, different people carry out experiments (or series of experiments) and execute slightly different manual actions, which can in turn lead to different results. Unfortunately, very often it only becomes apparent after the fact that series of experiments have to be repeated due to missing metadata. At one specific experiment in the composite peening project, for example, it was revealed only after the test series had been evaluated that a heat treatment failed due to incomplete temperature measurement. Table 3 puts this into monetary terms.

**Table 3:** The approximate costs of a test series in composite peening.

| STEP | WORKING TIME | COSTS |
|---|---|---|
| **Sample preparation** | Research assistant (40 h) | 610 € |
|  | Technician (8 h) | 260 € |
| **Prearrangement of the experiment and experimental procedure** | Research associate (5 h) | 220 € |
|  | Research assistant (20 h) | 310 € |
| **Evaluation and presentation** | Research associate (8 h) | 350 € |

Taking Table 3 into account, a test series costs approximately 1,750 €. Assuming three test series would need to be unnecessarily repeated over the course of a PhD thesis, this results in a total savings potential of **5,250 €**.

### 2.3. Phase 3: FAIR Data in Data Analysis

For the evaluation of test series with new materials, common standard evaluations (e.g., DIN or ASME standards) have to be modified regularly, or completely new evaluation procedures have to be generated. Independently programmed evaluation scripts usually create reproducibility within the test series of one specific scientist, but for testing purposes sometimes several similar scripts are coexistent. FAIR documentation of the evaluation scripts supports the tagging of raw and processed data and helps to minimize errors. Much more critical is the usage of the script by other researchers without a suitable documentation. In the case of a personnel change, for example, the evaluation must be re-programmed from scratch if the experimental setup is slightly modified (different resolution rate, new sensor, new software). With a time expenditure of ten hours per evaluation (e.g., tensile test: DIN EN ISO 6892-1, ASTM E8), this results in a savings potential of **440 €** per evaluation. Assuming that during a PhD project several tests (bending tests, tribological tests, fatigue tests, topology evaluation) are to be evaluated, which go beyond standard tests, **2,200 €** can be saved in the scope of, conservatively speaking, five experiments.

### 2.4. Phase 4: FAIR Data in Publishing

As indicated above for Phase 1, the reusability of experimental data is limited without complete metadata. Furthermore, even with a complete description in the experimental section of a scientific publication or PhD project, a time-consuming effort is required to make graphically-represented data usable again. The simultaneous publication of data represented in a diagram as raw data and metadata allows for an exponential increase in reusability and thus the overall benefit of the scientific study; first for the domain in question and ultimately for the taxpayers, who in most cases funded the research.

The transfer of experimental data and simulation scripts to subsequent graduate students is critical for performing research at the highest possible level. Much (if not all) information obtained that has not been published is lost due to non-FAIR data and must be searched for in old directories or, in the worst case, has to be repeated. This leads to high costs and negative environmental effects due to the repetition of entire series of experiments.

### 2.5. Calculation of the Potential Savings

Table 4 lists the savings potential when applying the FAIR data concept in the scope of this specific PhD project. This results in an estimated saving potential of approximately **2,600 €** per year.

**Table 4:** Summary of the **conservative** calculation

| Literature review | € 350 |
|---|---|
| Lab work | € 5,250 |
| Data analysis | € 2,200 |
| **Total costs** | **€ 7,800** |

### 3. Conclusions

Using FAIR data, as just shown, can save considerable amount of money even in a PhD project in the field of engineering. The scope and thus the benefit of applying the FAIR data concept may vary for other projects, but it gives a first practical impression of the potential savings that FAIR data can generate. In the case of several projects and students, the benefits can at least be proportionately multiplied.

As an example, in the case of a research group consisting of five PhD students, € 10,000 per year would be available to upgrade laboratories towards FAIR data generation. In this sense, investments in FAIR data are able to repay themselves very quickly. With conservative upgrade costs of € 50,000 for new hardware (server, electricity), software (which is often open source), and education, the breakeven point would be reached after five years – approximately after two generations of PhD students.

According to DESTATIS 35,800 doctoral students are registered in engineering programs throughout Germany [12]. Loosely extrapolating this number, an increase in efficiency of € 71,000,000 per annum would be achievable for research in the engineering sciences through a holistic implementation of FAIR data. This is recognized by public funding agencies and in Germany, for example, where the National Research Data Infrastructure (NFDI) allocates around 70,000,000 € per year for all research sectors, out of which around € 11,500,000 per year are for engineering sciences [13].

While this study does have a focus on purely monetary aspects, what might be even more important is the aspect that an individual researcher's creativity might be inspired completely differently when they have much easier and less cumbersome access to a wider range of information and meta data.

## 4. Limitations and Outlook

This case study accounts for the immediately visible costs in a PhD research program. There are multiple additional time-saving opportunities that FAIR data offers, such as the option of applying large-scale automated analysis techniques, better resource management, or increased internal findability of results. Furthermore, FAIR data may also reveal trends that were previously overlooked due to a lack of a holistic view. Of course, there are costs involved when it comes to transitioning an individual principal investigator's lab to FAIR practices. These costs by no means should be taken out of the equation. Scientific communities and funding agencies need to discuss how not only one-time investments in, for example, data storage infrastructure can be financed, but – and perhaps even more importantly – the long terms costs, e.g. for maintaining software for the implementation of FAIR practices, need to be covered.

Tackling global challenges by enabling the automated connection between currently disconnected scientific domains is one of the most prominent strategies that we have at hand. FAIR data is one of the first practical steps in that direction, and if successful, today's investment and immediate savings quoted above most likely will look like rounding errors for future generations of researchers.

Last but certainly not least, having FAIR data implemented internally makes it easier to *publish* FAIR data. Firstly, many modern challenges require combined efforts of many scientists, and exchanging data via email or within managed consortia is not practical. Secondly, sharing data can be viewed as scientifically ethical and fair, when tax-based funds are used for research: the public and interested researchers should be able to access what are now often hidden aspects of scientific results.

**Contacts**
Michael Seitz – ORCID: 0000-0002-6750-4719 - email: Michael.Seitz@kit.edu
Nick Garabedian – ORCID: 0000-0003-4049-4212 - email: Nikolay.Garabedian@kit.edu
Ilia Bagov – ORCID: 0000-0002-9094-8959 - email: Ilia.Bagov@kit.edu
Christian Greiner – ORCID: 0000-0001-8079-336X - email: Christian.Greiner@kit.edu